\documentclass[usenatbib]{mn2e}
\usepackage{changes}
\usepackage{graphicx}

\usepackage{amssymb}
\usepackage{color}
\usepackage{float}
\usepackage{amsmath}
\usepackage[title]{appendix}
\usepackage{booktabs}
\usepackage{multirow}
\usepackage{siunitx}
\usepackage{indentfirst}
\usepackage{physics}

\newcommand{\bn}{\begin{enumerate}}
\newcommand{\en}{\end{enumerate}}
\newcommand{\bi}{\begin{itemize}}
\newcommand{\ei}{\end{itemize}}

\def\gtorder{\mathrel{\raise.3ex\hbox{$>$}\mkern-14mu
    \lower0.6ex\hbox{$\sim$}}}
\def\ltorder{\mathrel{\raise.3ex\hbox{$<$}\mkern-14mu
    \lower0.6ex\hbox{$\sim$}}}

\setcitestyle{aysep={},citesep={,}} 

\title[Leading Spiral Arms]{Bar-driven Leading Spiral Arms in a Counter-rotating Dark Matter Halo}

\author[E. Lieb, A. Collier, and A.~M. Madigan]
{Emma Lieb$^{1}$, 
Angela Collier$^{1}$\thanks{E-mail: angela.collier@colorado.edu},
and Ann-Marie Madigan$^{1}$
\\
\footnotemark    
$^{1}$ JILA and Department of Astrophysical and Planetary Sciences, CU Boulder, Boulder, CO 80309, USA}
\begin{document}

\date{Accepted ?; Received ??; in original form ???}


\maketitle

\begin{abstract}
An overwhelming majority of galactic spiral arms trail with respect to the rotation of the galaxy, though a small sample of leading spiral arms has been observed. The formation of these leading spirals is not well understood. 
Here we show, using collisionless $N$-body simulations, that a barred disc galaxy in a retrograde dark matter halo can produce long-lived ($\sim3$ Gyr) leading spiral arms. Due to the strong resonant coupling of the disc to the halo, the bar slows rapidly and spiral perturbations are forced ahead of the bar. We predict that such a system, if observed, will also host a dark matter wake oriented perpendicular to the stellar bar. 
More generally, we propose that any mechanism that rapidly decelerates the stellar bar will allow leading spiral arms to flourish. 

\end{abstract}



\section{Introduction}
\label{sec:intro}

The formation and evolution of galactic spiral arms is a fundamental problem in astronomy. Nearly one hundred years of theoretical and observational work (e.g., \citet{hub26}, \citet{lind63}, \citet{linshu64}, \citet{elmgreen&elmgreen82}, \citet{sell-carl1984} and many more) have not yet resulted in a complete dynamical description of the mechanisms that generate, and regenerate, these structures. Recent research show renewed interest in questions first asked in the early days of the field \citep{hart17,peter19,yu19,pringle19,mata19,pet2020,sellwood&gerhard20}. In particular, are spiral arms material structures or density waves? 

Similarly, we re-open an old question, are long-lived spiral arms allowed to lead? Spiral arms are called trailing if the tails of the arms point in the opposite direction of galactic rotation, and leading if the tails point in the same direction.  This subject was once an active discussion \citep{lb41,irwin52,evans56}. However interest waned when a survey by \citet{deVac58} found no evidence of leading arms in their sample.

Decades later, \citet{pashasmirnov82} and \citet{pasha85} led systematic searches for leading spirals. A conclusive detection of leading arms in a galaxy requires observing the galaxy at a specific orientation; one must see the galaxy sufficiently face-on to view the spiral structure, but also at such an inclination that allows the measurement of the rotation direction. The observer must also know the side of the galaxy that is nearest to our own \citep{b&t}.  From their sample of nearly two hundred, \citet{pashasmirnov82} found only four disc galaxies with evidence of leading spirals.  Two of those cases have since been re-evaluated and found to host trailing arms \citep{sharp1985}. The remaining two (NGC 3786 and NGC 5426) have not been conclusively proven to be leading. 
 
More recent examples however confirm leading spirals in galaxies, NGC 4622 \citep[][]{buta92,buta05,buta08} and ESO 297-27 \citep{grouchey08}, with each galaxy hosting multiple sets of spiral arms that open in opposite directions---requiring one set to be leading and one set to be trailing. The only convincing example of a singular pair of leading arms is in galaxy IRAS 18293-3134 \citep{vaisanen08}. 
Galaxy morphology may provide clues as to the formation mechanism behind the observed leading arms. IRAS 18293-3134 is a gas-rich luminous infrared spiral galaxy; the leading spirals may be tidally induced by a smaller elliptical companion. ESO 297-27 and NGC 4622 are ringed, unbarred spiral galaxies. The dynamics of such galaxies are still not well understood but nuclear rings are shown to be a result of bar evolution in simulations (e.g., \citet{buta96}.)  As a result, no one mechanism of leading spiral formation can be ruled out by the small number of observed systems. Currently, the expectation is that leading spirals can be dynamic in nature, bar-driven, or the result of tidal interactions. We introduce a formation scenario for bar-driven leading spirals in this work.

 The low number of observed leading arms hints at the rarity of the phenomenon. However any detection of leading arms places interesting constraints on spiral arm theory. Density wave models of spiral structure allow for leading waves (see review by \citet{shu16}), however trailing waves are should be stronger due to amplification by galactic shear \citep{toomre81}. Theoretical explorations of tidally-induced spiral arms suggest that retrograde companions can produce long-lived, leading single ($m=1$) spirals in the disc \citep{athan78,thom89}. These results do not conclusively explain the observed leading spirals which are $m=2$ or higher multiples.  However, it highlights the importance of retrograde kinematics.

This paper explores the formation and regeneration of leading spiral arms in barred disc galaxies. More specifically, we study the important effects of retrograde dark matter orbits on the morphology and evolution of barred disc galaxies. This paper is a continuation in a series studying the effect of counter-rotation on stellar and dark matter morphological and kinematic evolution. Stellar bar evolution in the presence of retrograde spinning halos of varying degrees is analyzed in \citet{coll19}. The authors find that while retrograde rotation in a halo delays the bar instability it does not limit bar growth. Halo-bar coupling has also been shown to strongly influence bar dynamics \citep{CollierandMad20}; retrograde dark matter particles sink into a dark matter wake oriented perpendicular to the stellar bar, the formation of which results in a strong negative torque and a rapid slow-down of the stellar bar. 
In this paper we discuss a feature of counter-rotating halos that has not yet been explored---long-lived leading spiral arms.

\begin{figure}
\centerline{
 \includegraphics[width=0.5\textwidth]{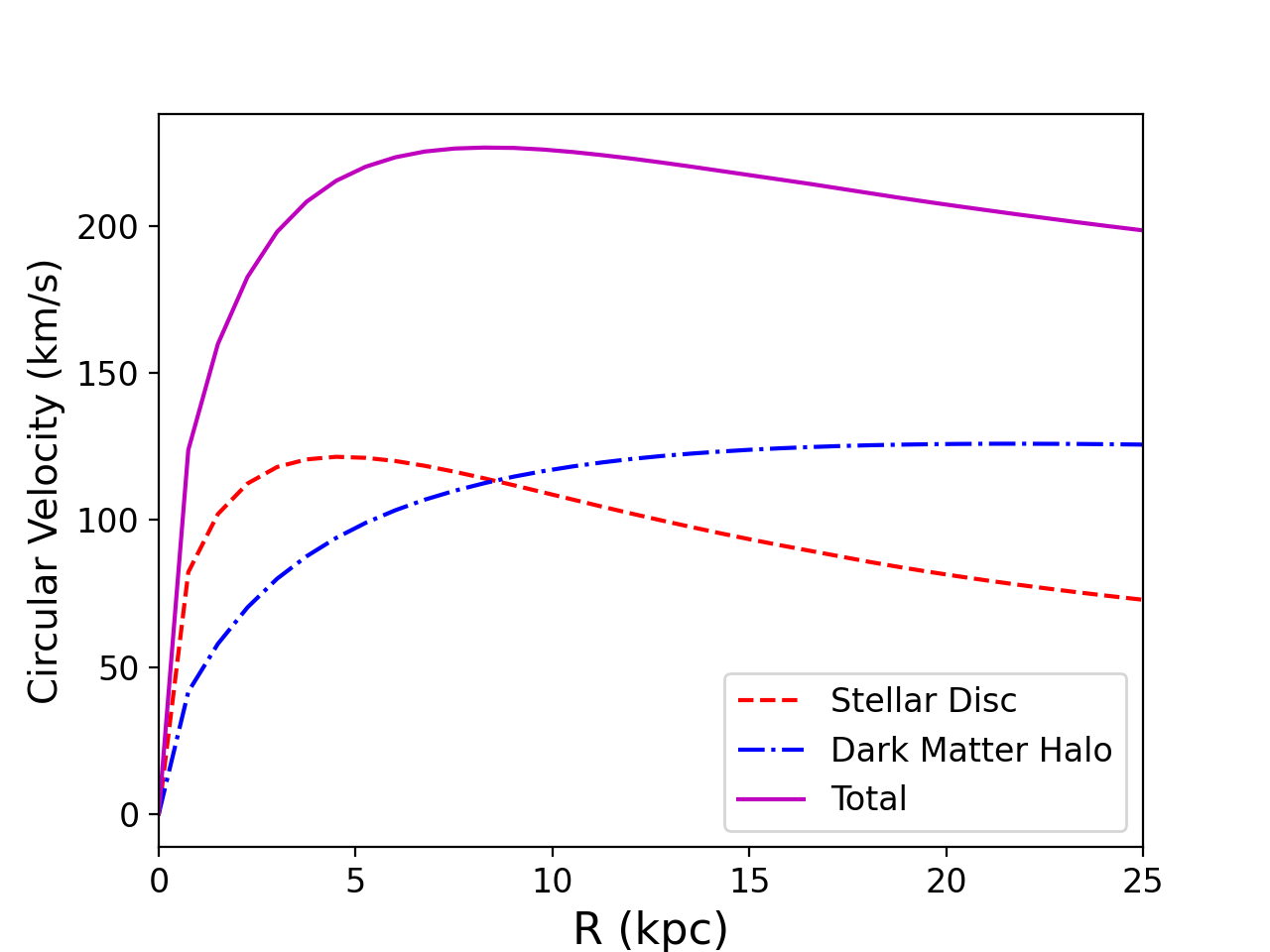}}
\caption{The galactic velocity curve (solid magenta) of our simulation at $t=0$, with the stellar disc (red dashed) and dark matter halo (blue dash-dotted) components.}
\label{fig:ic}
\end{figure}

\begin{figure*}
\centerline{
 \includegraphics[width=\textwidth]{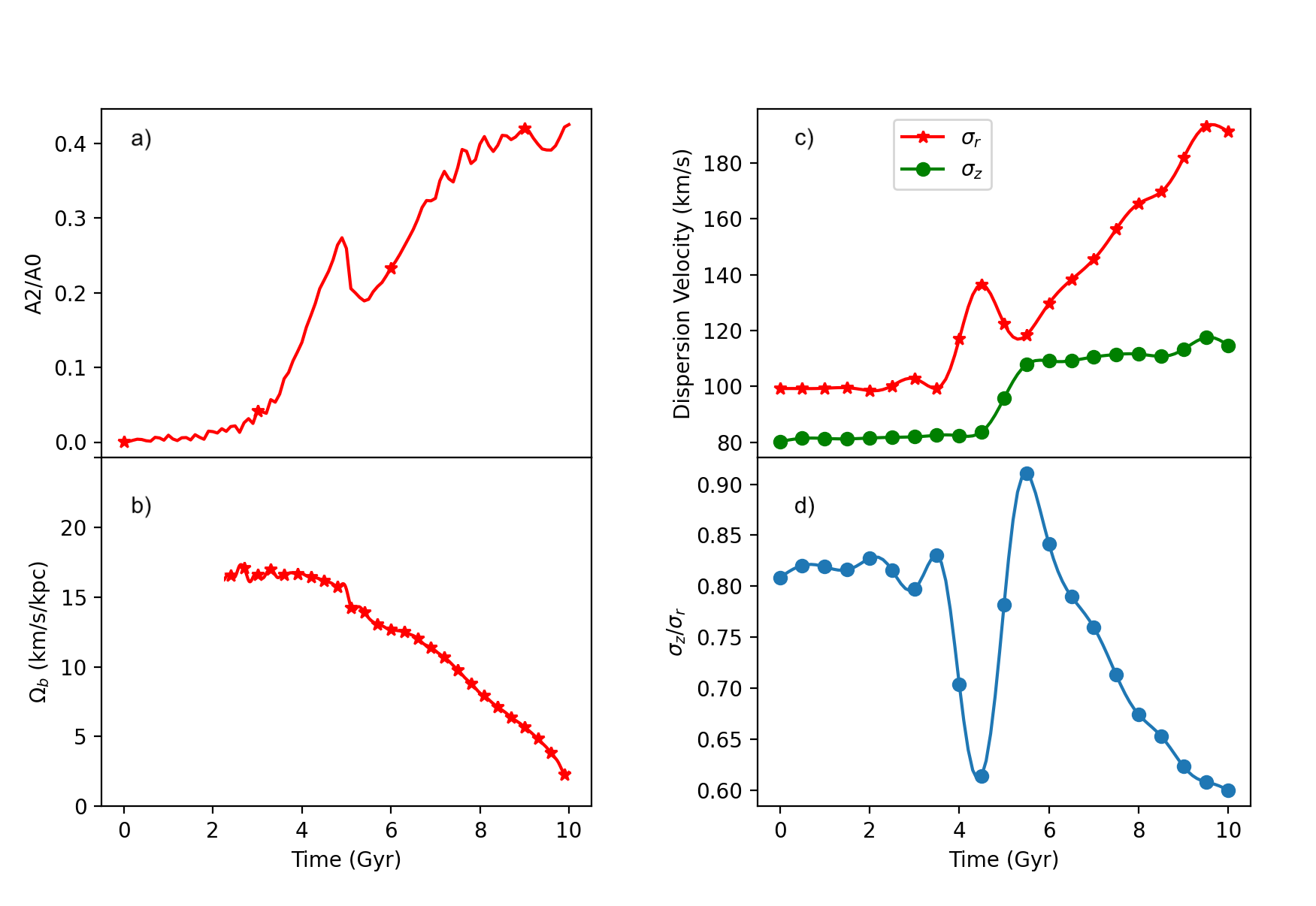}}
\caption{Time evolution of bar parameters. a: bar strength ($A_2/A_0$) in the $x/y$-plane which compares the growth of the Fourier $m=2$ mode to $m=0$ mode.  b: bar pattern speed ($\Omega_b$) in km/s/kpc. c: time evolution of the radial ($\sigma_r$) and vertical ($\sigma_z$) dispersion velocities for the stellar disc. d: ratio of $\sigma_z/\sigma_r$. The minima in the ratio are predict when this particular disc becomes unstable to bending modes.}
\label{fig:a2}
\end{figure*}

\section{Numerical Simulation}
\label{ics}

We simulate a disc embedded in a live dark matter halo with retrograde rotation (with respect to the stellar disc). The halo density is initialized with an NFW-inspired \citep{nava96} profile,

\begin{equation}
\rho_{\rm h}(r) = \frac{\rho_{\rm s}\,e^{-(r/r_{\rm t})^2}}{[(r+r_{\rm c})/r_{\rm s}](1+r/r_{\rm s})^2},
\end{equation}
where $\rho(r)$ is the dark matter  density in spherical coordinates, $\rho_{\rm s}$ is the fitting density parameter ($\approx7$ $10^{10}\,M_\odot$/kpc$^3$), $r_{\rm s}=9$\,kpc is the characteristic radius, and $r_{\rm c}=1.4$\,kpc is a central density core.   The Gaussian cutoff is applied at $r_{\rm t}=86$\,kpc. The dark matter  halo contains $7.2\times 10^6$ particles and the halo mass is $M_{\rm h} = 6.3\times 10^{11}\,M_\odot$.

The halo velocities are found by using an iterative method from \citet{rodio06}, see also \citet{rodio09}. The halo is initialized with the desired density distribution with particle velocities set to zero. The system is then evolved for a short time (0.3 Gyr).
 Each particle in the initial halo is assigned a velocity magnitude from the evolved system using a nearest neighbors scheme. The direction of these velocities are randomized.
 This constitutes one iteration. Iterations proceed until the initial velocity distribution is indistinguishable from the evolved velocity distribution. This creates a halo with a non-rotating velocity distribution. Cosmological halos however are generally spinning.  The cosmological spin parameter is $\lambda \equiv J_h/\sqrt{2}M_{\rm vir}R_{\rm vir}v_c$ where $J_h$ is the total angular momentum of the dark matter halo, $M_{\rm vir}$ and $R_{\rm vir}$ are the viral mass and radius of the dark matter halo, and $v_c$ is the circular velocity of the system at $R_{\rm vir}$. Cosmological simulations have found that the range of spin parameters can be fit by a lognormal distribution,

\begin{equation}
P(\lambda) = \frac{1}{\lambda (2\pi\sigma_{\lambda})^{1/2}} \textrm{exp} \bigg[-\frac{\textrm{ln}^2 (\lambda/\lambda_0)}{2\sigma_{\lambda}^2}\bigg],
\end{equation}
where $\lambda_0=0.035\pm 0.005$ and $\sigma_{\lambda}=0.5\pm 0.03$ are the fitting parameters \citep[][]{bull01,hetznecker}.

To create a halo that spins retrograde with respect to the disc we reverse the tangential velocities of all prograde halo particles. The new velocity distribution maintains the solution to the Boltzmann equation and does not alter the velocity profile \citep{lynd60,wein85}, so the equilibrium state is preserved. The fully retrograde halo is within the expected value of halo spin found in cosmological simulations ($\lambda=0.101$).

 The volume density of the exponential stellar disc is

\begin{eqnarray}
\rho_{\rm d}(R,z) = \bigl(\frac{M_{\rm d}}{4\pi h^2 z_0}\bigr)\,{\rm exp}(-R/h)
     \,{\rm sech}^2\bigl(\frac{z}{z_0}\bigr),
     \label{eqn:exp_disc}
\end{eqnarray}
where $M_{\rm d}$ is the disc mass, $h=2.85$\,kpc is its radial scale length, and $z_0=0.6$\,kpc is the vertical scale height.  The stellar disc has $N_{\rm d} = 0.8\times 10^6$ particles and the disc mass is $M_{\rm d} = 6.3\times 10^{10}\,M_\odot$. The radial and vertical dispersion velocities of stellar particles are given by

\begin{equation}
    \sigma_R(R) = \sigma_{R,0} e^{-R/2h}
\end{equation}
\begin{equation}
    \sigma_z(R) = \sigma_{z,0} e^{-R/2h}
\end{equation}

where $\sigma_{R,0}$ = 100 km/s and $\sigma_{z,0}$ = 80 km/s. The initial potential of each component is shown by the velocity curve plotted in Figure \ref{fig:ic}. 
We evolve the simulation using the $N$-body part of the tree-particle-mesh Smoothed Particle Hydrodynamics (SPH/$N$-body) code GIZMO \citep{hop15}.  Our code units for mass, distance, and time are  $10^{10}\,M_\odot$, 1\,kpc, and 1\, Gyr.

\section{Results}
\label{results}

\subsection{Stellar Bar Evolution and Formation of the Dark Matter Wake}

The disc in our simulation is nearly axisymmetric at $t=0$ Gyr and then forms a bar around $\sim 3.5$ Gyr. The bar instability is delayed by the counter-rotation of the halo. \citep{Saha13, coll18} The evolution of stellar bar strength is shown in Figure \ref{fig:a2}a. The strength of the bar is defined by the ratio of the Fourier $m=2$ mode to the $m=0$ mode,

\begin{eqnarray}
\frac{A_2}{A_0} = \frac{1}{A_0}\sum_{j=1}^{N_{\rm d}} m_{\rm j}\,e^{2i\phi_{\rm j}},
\end{eqnarray} 
which is calculated by summing over all disc particles with $R \leq 14$ kpc, and mass m = $m_j$ at azimuthal angle $\phi_j = \tan^{-1}(y/x)$. 

Stellar bars increase the fraction of radial orbits in the disc which in turn increases the radial velocity dispersion. This makes the disc increasingly unstable to bending modes (the buckling instability) in the $x/z$-plane. \citep{raha91,merritt&sellwood94,valpuesta04,coll20,sellwood&gerhard20}. The buckling instability can be observed in the brief drop in $A_2/A_0$ at $\sim 5.5$ Gyr. After which the bar continues to grow in length and strength. The buckling instability can occur multiple times during the lifetime of the bar as seen in our simulation \citep{valpuest06}.
A second buckling-like instability occurs in the last Gyr of the simulation. To quantify the timescale of the buckling instabilities, we plot the radial ($\sigma_r$) and vertical ($\sigma_z$) dispersion in Figure \ref{fig:a2}c measured in a $1$ kpc bin around $R=3$ kpc. Figure \ref{fig:a2}d shows the ratio of $\sigma_z/\sigma_r$. The exact value at which discs become unstable to bending modes is dependent on the properties of the disc and halo \citep{hunttoom69,merritt&sellwood94,b&t}. However, the locations of the minima in Figure \ref{fig:a2}d at $\sigma_z/\sigma_r \approx 0.6$ match the times at which the buckling instabilities occur.  Buckling thickens the disc and increases vertical dispersion velocities moving the disc out of the unstable region. 

A stellar bar acts as a conduit of angular momentum transfer. As the stellar bar rotates it brakes against the outer disc and dark matter halo, loses angular momentum, and slows down (e.g. \citet{lynden-bell72}). We plot the pattern speed of the stellar bar in Figure \ref{fig:a2}b. The bar slows down until the first buckling instability. The bar at this point experiences a small, steep drop in pattern speed and then we see an increase in the rate of its slowdown for the remainder of the simulation. The rapid deceleration is due to the torque acting on the stellar bar from low inclination, retrograde dark matter halo orbits. \citet{CollierandMad20} show that angular momentum is carried away from the bar by retrograde halo orbits that are torqued so strongly that they reverse their orbital direction. Over secular times, these orbit reversals greatly reduce the pattern speed of the bar (Figure \ref{fig:a2}b). Remaining retrograde halo orbits form a dark matter wake that is oriented perpendicular to the more massive stellar bar.

\begin{figure}
\centerline{
 \includegraphics[width=0.5\textwidth]{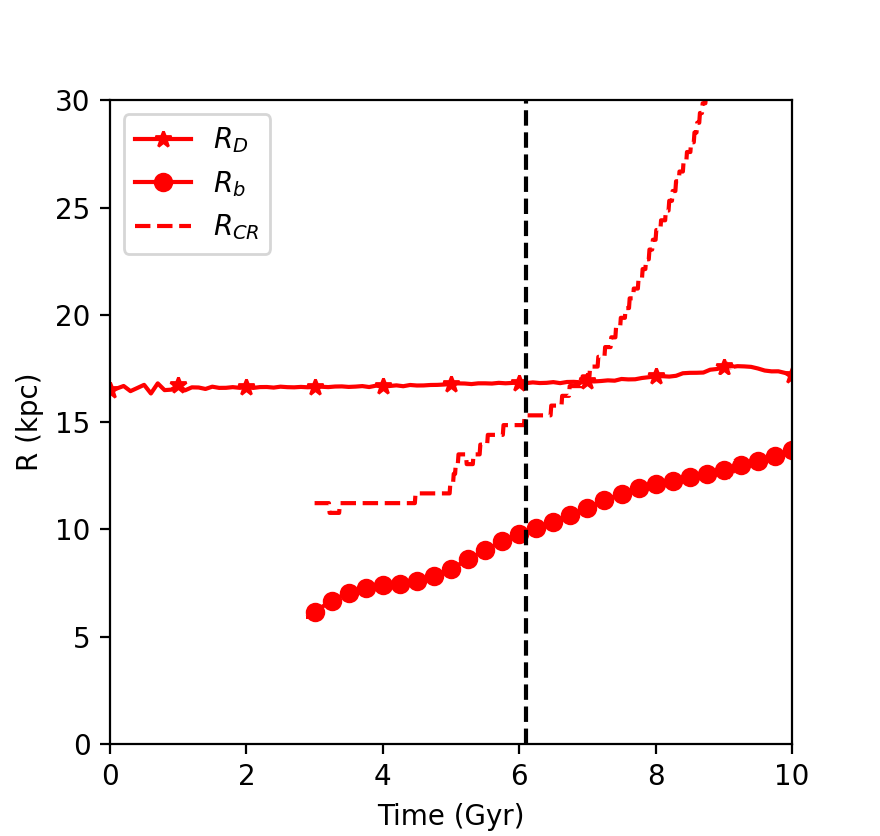}}
\caption{Evolution of stellar disc radius ($R_D$) in kpc (starred line), the bar radius ($R_b$) in kpc (dotted line) and the corotation ($R_{\rm{CR}}$) radius (red dashed line.) The black vertical dashed line represents the time when leading spirals emerge in the simulation.}
\label{fig:rcr}
\end{figure}

\begin{figure*}
    \centering
    \includegraphics[scale = .3]{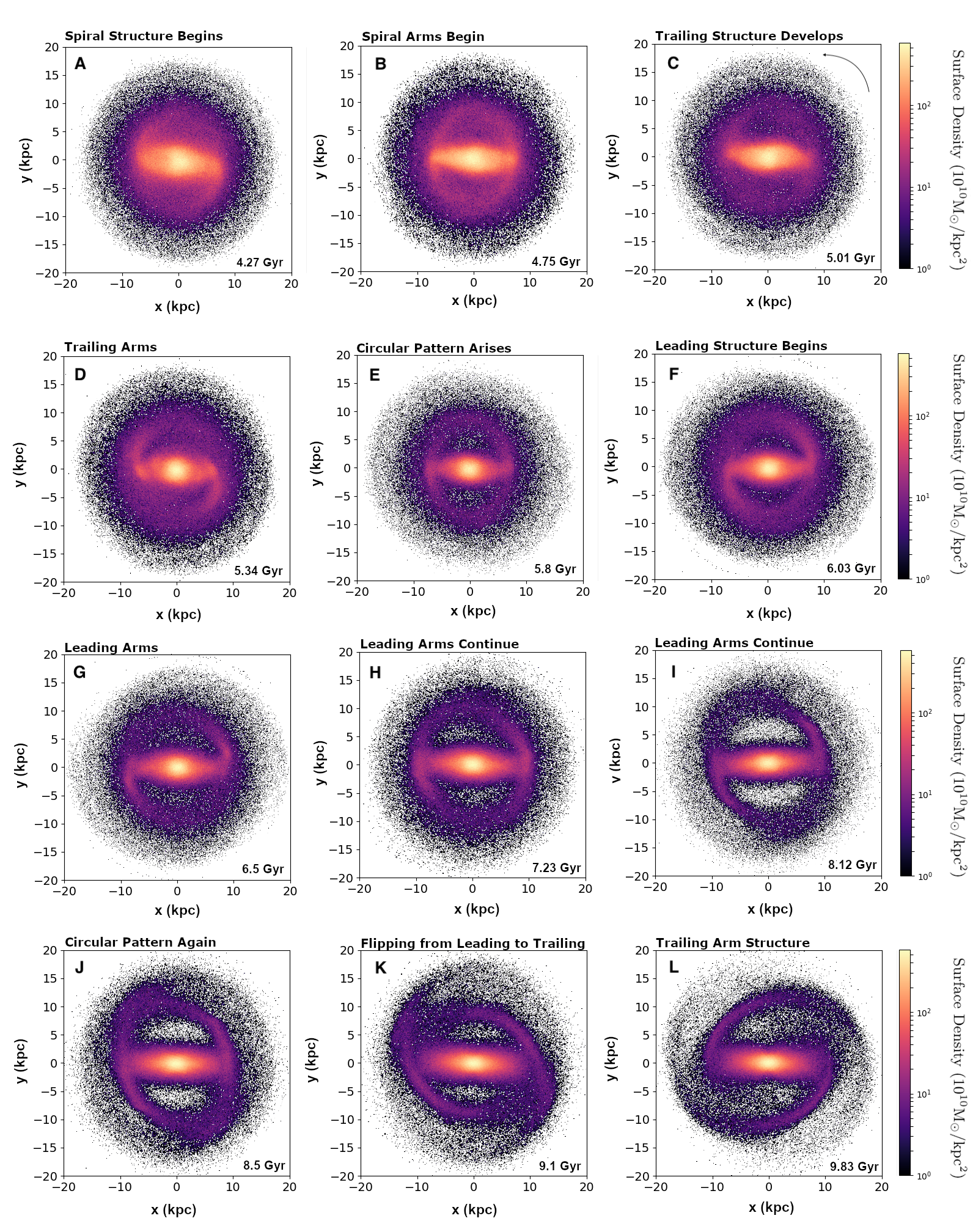}
    \caption{Face-on surface density profiles of the stellar disc at various times after bar formation. Time is indicated in the bottom right corner of each panel and the direction of the bar rotation is counter-clockwise as indicated by the arrow in panel C. Leading spiral arms emerge in panel F. The second buckling happens around the time of panel K allowing the leading spirals to detach from the bar and wind back around the disc and attach to the opposite end of the bar, appearing trailing in panel L.}
    \label{fig:spiral_dens}
\end{figure*}

\begin{figure}
    \centering
    \includegraphics[scale = 0.45]{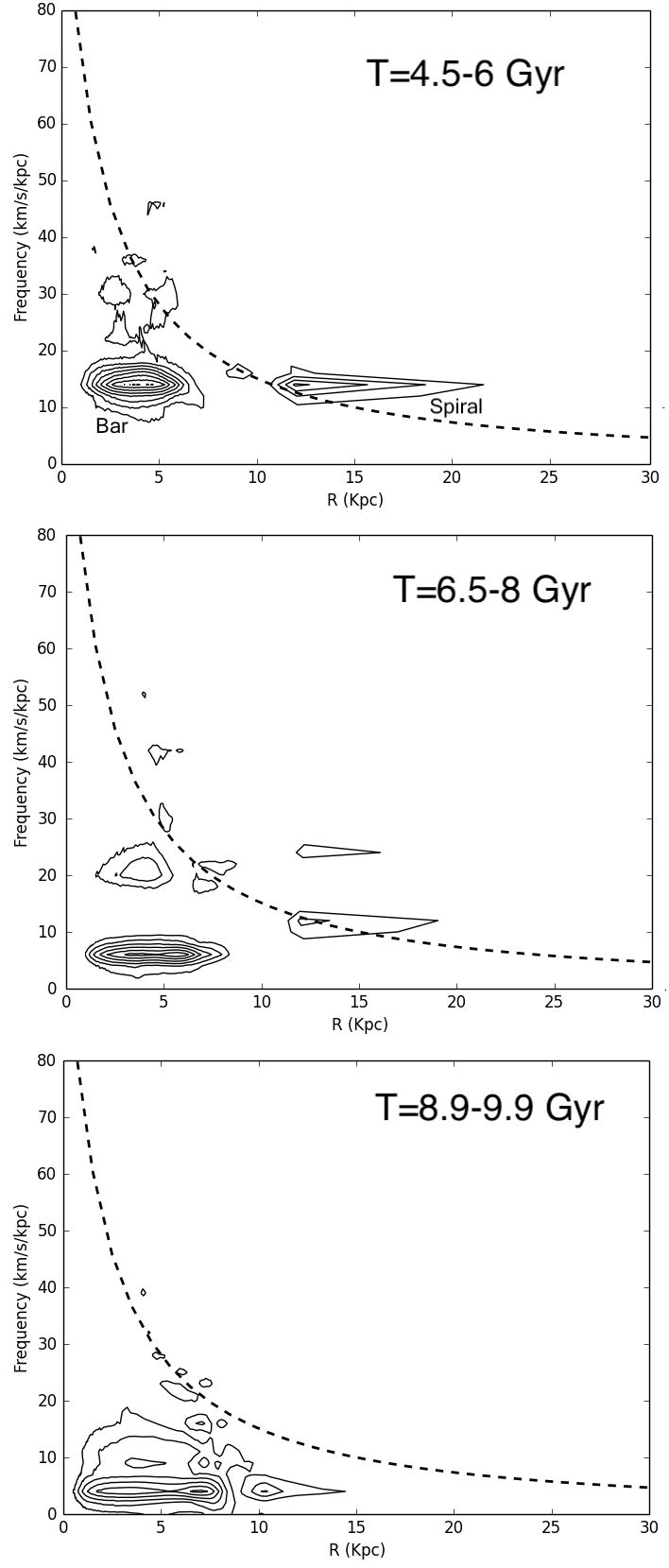}
    \caption{Contours of the power spectrum of frequency and radius for various time steps. Curves show the radial extent and pattern speeds of the bar and spiral patterns, as indicated by the text in the top panel. The dashed line denotes the angular velocity of the disc at $t = 0$.}
    \label{fig:freq}
\end{figure}

\subsection{Evolution of Corotation Resonance and Leading Spirals}

The position of corotation resonance depends on the pattern speed of the bar. In our simulation, halo-bar coupling rapidly slows the stellar bar moving the corotation radius beyond the stellar disc. We plot the time evolution of disc radius ($R_D$), bar length ($R_b$) and corotation radius ($R_{\rm CR}$) in Figure \ref{fig:rcr}. $R_D$ is the radius containing $97\%$ of the stellar material which is similar to the observable $R_{25}$,  the $25$th magnitude isophote of the galactic disc, the Holmberg radius. 
Bar length ($R_b$) is given by the radius where the ellipticity of isodensity contours of stellar particles in the $xy-$plane drops to $85\%$ its maximum value \citep{marti06,coll19}. Here we show that we have created an environment that allows the bar to decelerate rapidly, pushing $R_{\rm CR}$ beyond the disc radius in a short time scale. 

Bar-driven spiral arms form around 4 Gyr in the simulation. In Figure \ref{fig:spiral_dens}, we plot surface density maps of the face-on ($x/y$-plane) stellar disc at multiple time steps.
Trailing spiral arms are clearly present by 4.27 Gyr (panel A) and begin with trailing sense (panels B - D). The spirals then appear to wrap around the bar forming a complete loop until leading spiral arms emerge around 5.8 Gyr (panel E). These leading arms are long-lived, existing for almost 3 Gyr (panels F - J). At $\sim$ 9 Gyrs the bar undergoes a second buckling instability. The weakening of the bar results in the detachment of the leading spiral arms from the ends of the bar. The detached arms wind around the bar (as the pattern speed of the wave is larger than that of the bar) and re-attach to its opposite end (panels K, L) giving the appearance of trailing arms once more.

 Figure \ref{fig:freq} show contours of the power spectrum of frequency and radius at three times during the simulation. By averaging over many time steps, the pattern speeds of the bar and spiral patterns can be determined from the power spectrum. To create this figure we measure $A_2/A_0$ within $0.25$ kpc bins along the radius of the disc. We apply a Fourier transform to the time series of this data to create the power spectrum. Due to the rapid reduction in pattern speed of the bar and the length of the time series required to make this plot, the results are quite noisy. This technique was first used in \citet{sellwoodsparke88}. One of the findings of \citet{sellwoodsparke88} is that the bar and spirals may have different pattern speeds in frequency space but still appear connected in density space. Unique to our work, we find that for an extended period of time ($\sim2$ Gyr) the spiral modes are moving faster than the  bar. The fast moving spirals disconnect from the bar ends during the second buckling event and wrap around the disc.
 
 The top frame of Figure \ref{fig:freq} shows trailing spirals which are located outside the corotation radius. Swing amplification will amplify trailing spirals outside corotation. The middle frame shows spiral arms inside the corotation radius, traveling at a faster speed than the bar. Because they are traveling faster than the bar, they move ahead of the bar and appear to be leading. The spirals are bar-driven, but due to the buckling instability their regeneration is interrupted. This briefly allows the spirals to be moving a pattern speed different than the bar. Hence in this middle panel, the spirals appear faster than the bar. The final panel shows a very slow bar and equally slow spiral mode, regenerated from the bar.

 Next we discuss the movement of bar-driven angular momentum throughout the galaxy in the presence of long-lived leading spirals.

\subsection{Angular Momentum Flow}

\begin{figure*}
\centerline{
 \includegraphics[width=\textwidth]{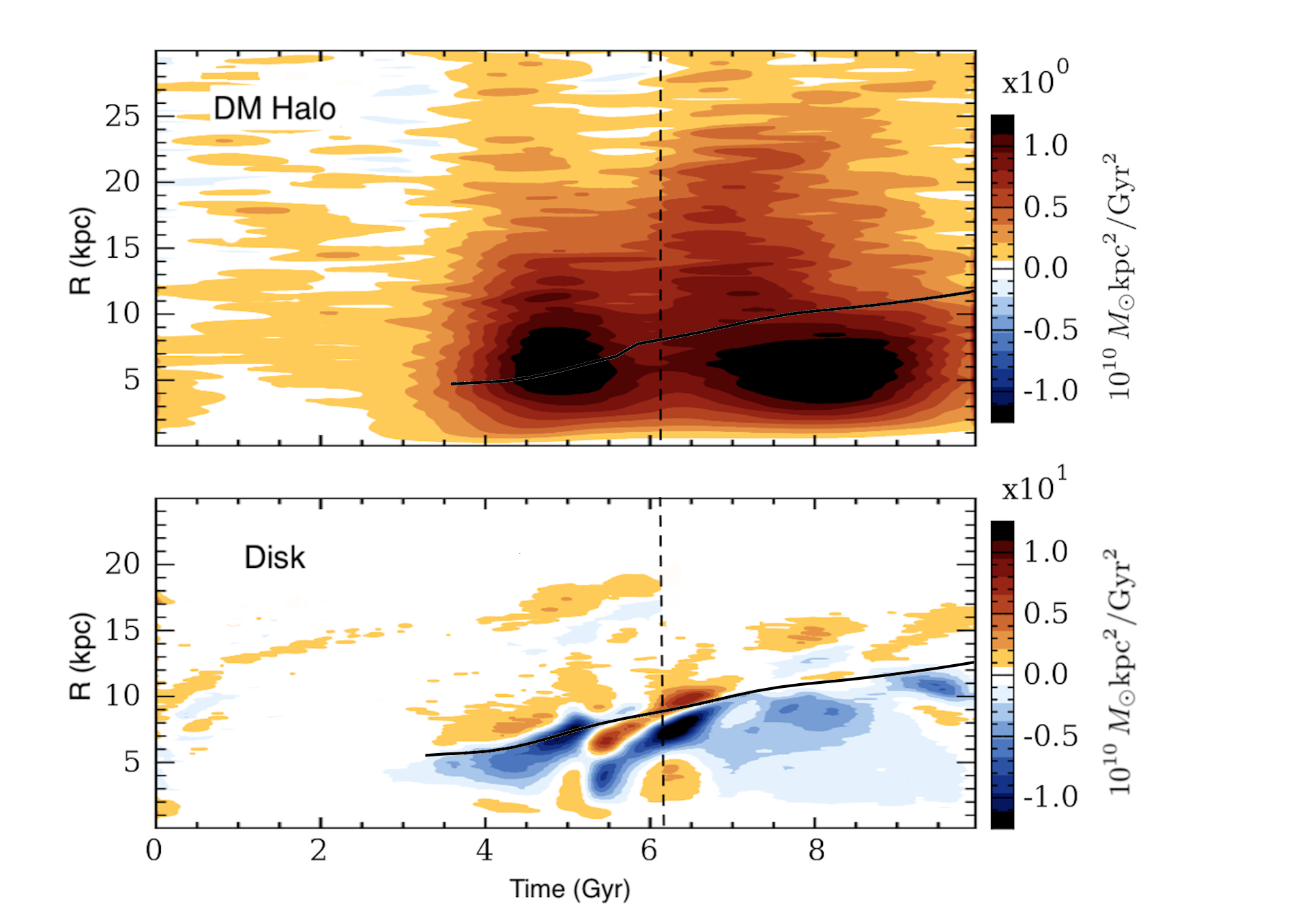}}
\caption{Torque maps for the dark matter halo (top) and stellar disc (bottom). The color palette shows blue for emission of angular momentum and red for absorption of angular momentum. Torque at each radius is plotted along the $y-$axis for each time step along the $x-$axis. The data is binned into $1$ kpc in R and limited to $|z|<10$ kpc. Solid black lines represent the length of the dark matter wake (top) and stellar bar (bottom).
Leading spiral arms emerge after the dashed line.}
\label{fig:angmo2}
\end{figure*}

To visualize the flow of angular momentum from stellar disc to halo, we create a map showing angular momentum movement throughout the simulation in Figure \ref{fig:angmo2}. We use the method first described in \citet{villa-vargas-09}. First, the disc and dark matter halo particles are binned into cylindrical shells of $1$ kpc. At each time step the total amount of angular momentum, $J$, is measured. We then plot a color map of the rate of change of $J$ (i.e., torque) averaged over 5 times steps ($0.01$ Gyrs each) in cylindrical shells as a function of R (kpc) along the $y-$axis and time (Gyr) along the $x-$axis. The color palette shows a gain (absorption) of angular momentum as red and a loss (emission) of angular momentum as blue.  Leading spirals emerge after the dashed line. The bar length for the stellar bar in the disc and the perpendicular dark matter bar in the halo are plotted as solid lines.

First, we look at the disc. The bar forms at $\sim4$ Gyr and is quickly followed by the buckling instability at $\sim5$ Gyr. 
The bar moves angular momentum outward with the inner 5 kpc losing angular momentum and the outer disc expanding (and gaining angular momentum). 
The absorption of of angular momentum in the very outer disc stops at $\sim 6.1$ Gyr when the spiral arm winding direction reverses. We attribute this to the outer Lindblad resonance moving beyond the disc. While the spiral arms are leading, the outer disc ($R > 12$ kpc) continues to gain small amounts of angular momentum but at lower radii than earlier times. Near the end of the simulation during the second buckling event ($\sim 9.1$ Gyr), the leading spiral arms wrap around the bar which is associated with a smooth continuous increase in angular momentum from $12-15$ kpc. 

Next, we look at the angular momentum flow in the dark matter halo. We first note that the scale on the color palette is an order of magnitude smaller than the disc. The dark matter halo in this simulation acts as a sink of angular momentum, showing deep absorption, especially in the inner 10 kpc of the halo. The first deep absorption feature is associated with the bar instability which results in the swift movement of angular momentum out of the inner disc. The second deep absorption feature is associated with the renewed strengthening of the stellar bar. The second buckling instability begins at $\sim9.1$ Gyrs. Buckling weakens the stellar bar \citep{marti06} at large radii reducing the torque on the dark matter halo.

An important feature of trailing spiral arms is the transport of angular momentum from the inner disc to larger radii. However, inside corotation leading arms can also move angular momentum outward \citep{lynden-bell72}. We find that angular momentum \textit{is} flowing out of the inner disc between $6-10$ Gyr and moving to both the outer disc and the dark matter halo.

\section{Discussion and Conclusion}
\label{conclusion}

We have investigated the formation of leading spiral arms in an isolated disc galaxy.
In our simulation, bar-driven perturbations in the disc move ahead of the bar as it decelerates forming leading spiral arms. One would expect leading arms to travel outwards towards corotation and be amplified and reflected as a trailing wave \citep{toomre81}. However the bar decelerates so rapidly that corotation is pushed beyond the disc and the leading spiral arms are maintained. The leading wave could also in theory be reflected by the edge of the disk if it is sufficiently sharp \citep{b&t}. However, the exponential density profile of our simulated stellar disk (Equation~\ref{eqn:exp_disc}) means it does not have a sufficiently sharp outer edge.

 Key to the formation of the leading arms is the decelerating bar. In our simulation this deceleration is a result of the galaxy being embedded in a counter-rotating (retrograde) dark matter halo. 
 A non-rotating halo hosts both a parallel dark matter bar made up of trapped prograde-orbiting halo particles and a perpendicular dark matter wake composed of retrograde-orbiting particles \citep{CollierandMad20}.
As the ratio of retrograde to prograde mass in a halo increases, the stellar bar is increasingly negatively torqued due to the large population of halo orbit reversals from retrograde to prograde. This causes rapid deceleration of the bar and pushes the corotation radius outward. In \citet{CollierandMad20}, we calculate that this happens for halo spin values of $\lambda \lesssim-0.06$. Though we have shown results from one extreme model, we have run additional simulations of counter-rotating halos ($\lambda = -0.06, -0.075, -0.09$) to verify this effect. We find that the more retrograde material in the halo the longer the leading spirals are present. 

Dark matter simulations find the fraction of halos with $\lvert \lambda \rvert > 0.06$ to be about $10.5\%$ \citep{bull01}. Halo spin and disc spin should generally align however \citep{bett09,hahn10}. We expect the fraction of halos with large retrograde rotation with respect to the disc to be quite small, which would explain why bar-driven leading spiral arms are rare. 

Cosmological studies predict stellar mass to halo mass relations that can vary greatly from the single model ($M_d/M_h = 0.1$) presented  here \citep[e.g.,][]{behroozi13}. A smaller stellar mass to halo mass ratio may require less retrograde halo rotation (i.e., a lower value of $\lambda$) to produce the same effects described here due to the increased mass available for orbit reversals and ultimately bar deceleration.  Observing such a system could reveal the underlying kinematics of the dark matter halo. Specifically, we would expect this halo to have a net retrograde rotation with respect to the disc and an over-dense wake of dark matter perpendicular to the stellar bar in the plane.

Our main finding is more general. We hypothesize that any barred galaxy with a rapidly decelerating bar, and therefore an expanding corotation radius, should host leading spiral arms. Again, this phenomena should be quite rare as it requires the bar to be strongly negatively torqued without the disc suffering from the same torque (i.e., the bar must slow down while the outer disc maintains its speed).  Previous results produced long-lived m=1 leading spirals via tidal interactions (\cite{athan78}, \cite{thom89}). Here we have shown that long-lived leading bar-driven spirals are also possible.

\section*{Acknowledgements}
Many thanks to Jerry Sellwood, Debra Elmegreen, and Bruce Elmegreen for helpful conversations.
This work was supported by a NASA Astrophysics Theory Program under grant NNX17AK44G.  
A.M. gratefully acknowledges support from the David and Lucile Packard Foundation.
We use the RMACC Summit supercomputer, which is supported by the National Science Foundation (awards ACI-1532235 and ACI-1532236), the University of Colorado Boulder, and Colorado State University. The Summit supercomputer is a joint effort of the University of Colorado Boulder and Colorado State University.

\section*{Data Availability}

The data underlying this article will be shared on reasonable request to the corresponding author.

\bibliographystyle{mn}
\bibliography{leading_spirals}

\appendix
\section{}
\begin{figure*}
\centerline{
 \includegraphics[width=0.75\textwidth]{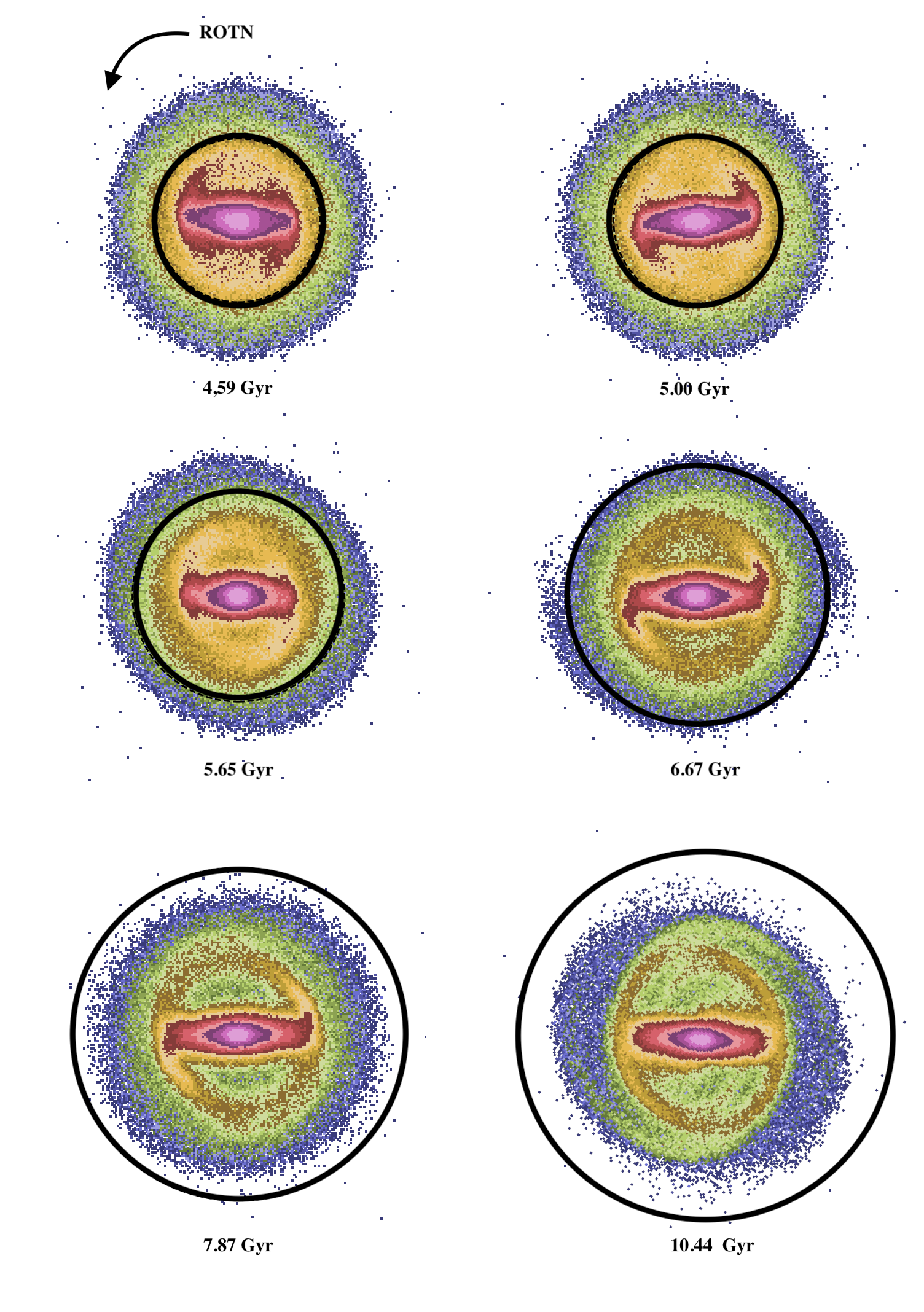}}
\caption{From dust to ashes? The fate of bar-driven leading spiral arms in a retrograde dark matter halo. Rotation direction is counter-clockwise. Corotation radius is shown by the thick black line.  We plot a frame every two rotations of the bar. Note the time between frames increases dramatically as the bar slows down. This plot is inspired by Figure 8 of \citet{toomre81}.}
\label{fig:app}
\end{figure*}

The fate of leading spiral wave packets is discussed in detail in the swing amplification paper by \citet{toomre81}. Their leading waves are amplified and reflected at corotation and then devoured by the inner Lindblad resonance. In contrast, our leading waves are continually generated by the stellar bar and maintained for long times as the corotation radius is pushed beyond the stellar disc. We recreate the aptly named `Dust to Ashes' (\citet{toomre81}, Figure 8) plot in Figure \ref{fig:app}. However, in our case the leading spirals do not die.

\end{document}